# Fractals in top non-leptonic decays

R.Odorico

University of Bologna, Department of Physics and Istituto Nazionale di Fisica Nucleare, Sezione di Bologna
Via Irnerio 46, I-40126 Bologna, Italy
e-mail: roberto.odorico@www.bo.infn.it

**Abstract**. A study of the use of fractals in top non-leptonic decays for the sake of discrimination against background is presented. Preliminary results show that fractals may provide a useful check for top event enrichment techniques.

## 1 Introduction

The lack of special features in top quark non-leptonic decays makes the extraction of this signal particularly hard to tackle with conventional techniques. The exploitation of decay particle correlations provides a possible way out to solve the problem. This is the case of neural-network classifiers, which have been used with some success [1-3]. Another tool may be provided by fractal geometry, which has already been applied to particle physics in other contexts, like e.g. in [4].

In this preliminary study we explore the capabilities of fractal geometry concepts in helping to discriminate between top quark non-leptonic decays and the underlying background. For this sake we concentrate on the $t\bar{t} \rightarrow$ non-leptonic channel in $p\bar{p}$ interactions at the energy of the Fermilab Tevatron. We look for fractal quantities which exhibit the most pronounced differences when calculated on $t\bar{t} \rightarrow$ non-leptonic and on background events. It turns out that for some of these quantities such differences are appreciable. They are not large enough to establish and identify by themselves a top signal, however they may provide useful checks for results obtained by other techniques.

## 2 Fractal dimensions

Fractal geometry is widely used in a variety of problems and does not need an introduction (for a recent particle physics application see, e.g., [4]). We only define entities which will be used in the present study. They are related to the concept of fractal dimension. In order to define it let us first consider simple examples based on uniform mass distributions.

If we consider a mass distribution over a line starting at $R = 0$, the mass $M(R)$ of the line segment up to $R$ has a behavior

$$M(R) \propto R^1 \qquad (1)$$

For a mass distribution uniform on a plane, one similarly has

$$M(R) \propto R^2 \qquad (2)$$

where $R$ now is the radius of a circle.

The exponent of R takes the name of fractal dimension. If we take the massive line and we consider it on a plane, the dependency $M(R) \propto R^1$ still holds with $R$ now representing the radius of a circle centered on the origin of the line. Thus its fractal dimension provides a geometrical characterization of the system.

For a mass distribution intermediary between the two considered, if it turns out that

$$M(R) \propto R^D \qquad (3)$$

$D$ having some non-integer value, we are still in the presence of a scaling law characteristic of a pure fractal structure. If we plot $log\ M(R)$ versus $log\ R$ we obtain a straight line whose slope is $D$.

If $D=D(R)$, i.e. $D$ depends on $R$, the distribution is said to represent a multifractal. Although a scaling law does not hold anymore, it may be still useful in such a case to employ the entities which are introduced for the handling of fractals.

In a particle physics application the mass is to be substituted with something else associated with the particles. In this study we consider particle weights given by the particle transverse energy $E_{T,i}$ or a power of it $E_{T,i}^q$. As to $R$ we define it as the particle pseudorapidity $\eta$ or as the particle azimuth $\phi$. More exactly we take

$$R = \eta/\eta_{max} \qquad (4)$$

or

$$R = |\phi|/\pi \qquad (5)$$

where $\eta_{max}$ is the maximum possible $\eta$, according to experimental acceptance, and $-\pi < \phi < \pi$ is measured with respect to the minor event axis $\vec{m}$ defined in the following section.

We define the quantity

$$I(R,q) = \frac{\sum_{\overline{R}<r<R} E_{T,i}^q}{\sum_{all} E_{T,i}^q} \qquad (6)$$

$\overline{R}$ in the numerator represents a lower cutoff which may turn out to be useful so as to exclude particle contributions from some $\eta$ or $\phi$ intervals. If $\overline{R} > 0$ it is in general $I < 1$.

One may define a R dependent fractal dimension by

$$D(R,q) = \frac{1}{q} \frac{\partial \log I(R,q)}{\partial \log R} \qquad (7)$$

The differentiation, however, leads to higher statistical errors. In the applications we consider in this study it is thus better to directly use the quantities $I(R,q)$ of eq. 6.

## 3 Event selection

$t\bar{t}$ and QCD background events are provided by the event generator COJETS 6.24 [5]. For $p\bar{p}$ interactions at a c.m. energy $E_{cm}$ = 1.8 TeV and a top quark mass $m_t$ = 175 GeV it yields a total $t\bar{t}$ cross-section $\sigma(t\bar{t})$ = 7.31 pb.

A series of cuts is imposed in order to: i) suppress the contribution of leptonic and semi-leptonic decays of the top quark; ii) reduce the amount of the background. Where possible, the cuts mimic those used in the neural-network D0 analysis [1-3].

For the suppression of the top leptonic and semi-leptonic decay contributions, events containing an electron or muon with a transverse momentum $p_T$ > 20 GeV or a missing transverse energy $E_T^{miss}$ > 20 GeV are excluded.

For the sake of background reduction the following requirements are imposed:

i) Total transverse energy $E_T^{tot}$ > 120 GeV;
ii) At least one muon with transverse momentum $p_T$ > 4 GeV and absolute pseudorapidity $|\eta|$ < 2.5.
iii) A linear aplanarity $A_L$ > 0.5. Considering only particles with $|\eta|$ < 2.5 (neutrinos excluded), after finding the thrust, or major axis $\vec{M}$, in the transverse energy plane, and defining the minor axis $\vec{m}$ as the axis orthogonal to it, linear aplanarity is defined as $A_L = \sum_i |\vec{E}_i \cdot \vec{m}| / \sum_i |\vec{E}_i \cdot \vec{M}|$, where $\vec{E}_i$ is the transverse energy vector of particle $i$.

For the integrated luminosity $L$ = 110.2 pb$^{-1}$ applying to the D0 event sample, the events left after all the cuts are: i) 128 events for $t\bar{t}$ ; ii) 3868 events for the background.

For the present study we employ 1000 events for $t\bar{t}$ and 1000 events for the background.

## 4 Fractal study for top non-leptonic decays

We have looked for $I(R,q)$'s which exhibit the most pronounced differences between $t\bar{t}$ → non-leptonic and background events. We have considered $R$ 's defined by eq.s 4 and 5. In eq. 6 we have set $\log \overline{R}$ = -½, which eliminates $\eta$ and $\phi$ regions where $I(R,q)$'s are negligible. A range of integer values has been considered for $q$.

## 5 Results

Fig.s 1-4 show a selection of the results of this preliminary study. From them it appears that some of the fractal quantities $I(R,q)$'s exhibit appreciable differences between $t\bar{t}$ → non-leptonic and background events.

Comparisons between fig.s 1,3 and 2,4 show that differences are more pronounced for $\phi$ rather than for $\eta$.

Comparisons between fig.s 1,2 and 3,4 show that differences become considerably higher for high $E_T$ particles.

**5 Conclusions**

From this preliminary study, one can conclude that fractal differences between $t\bar{t}$ → non-leptonic and background events may help to discriminate between the two channels. They do not appear to be strong enough to allow by themselves to establish or to identify a signal, however they may represent a useful check for other signal enrichment techniques like, e.g., neural networks [1-3].

**References**
1. B. Abbott et al. (D0 Collaboration), Phys. Rev. D **60**, 12001 (1999).
2. B. Abbott et al. (D0 Collaboration), Phys. Rev. Lett. **83**, 1908 (1999).
3. R. Odorico, Europhys. Lett. to be published.
4. L. A. Anchordoqui et al., Nucl. Phys. Suppl. **97**, 139 (2001).
5. R. Odorico, Comput. Phys. Commun. **72**, 238 (1992); updated program available in: http://www.bo.infn.it/preprint/odorico.html.

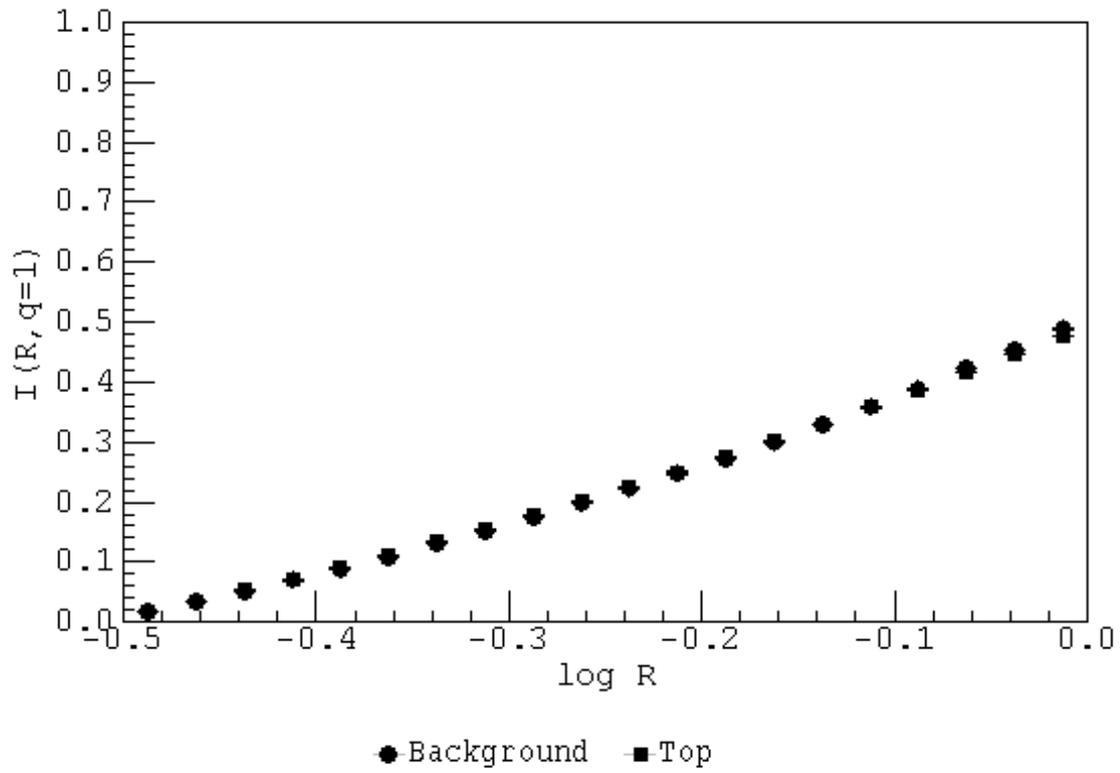

**Fig 1**. Comparison of *I(R,q=1)*'s, with *R* defined as in eq. 4, for $t\bar{t} \to$ non-leptonic and background channels.

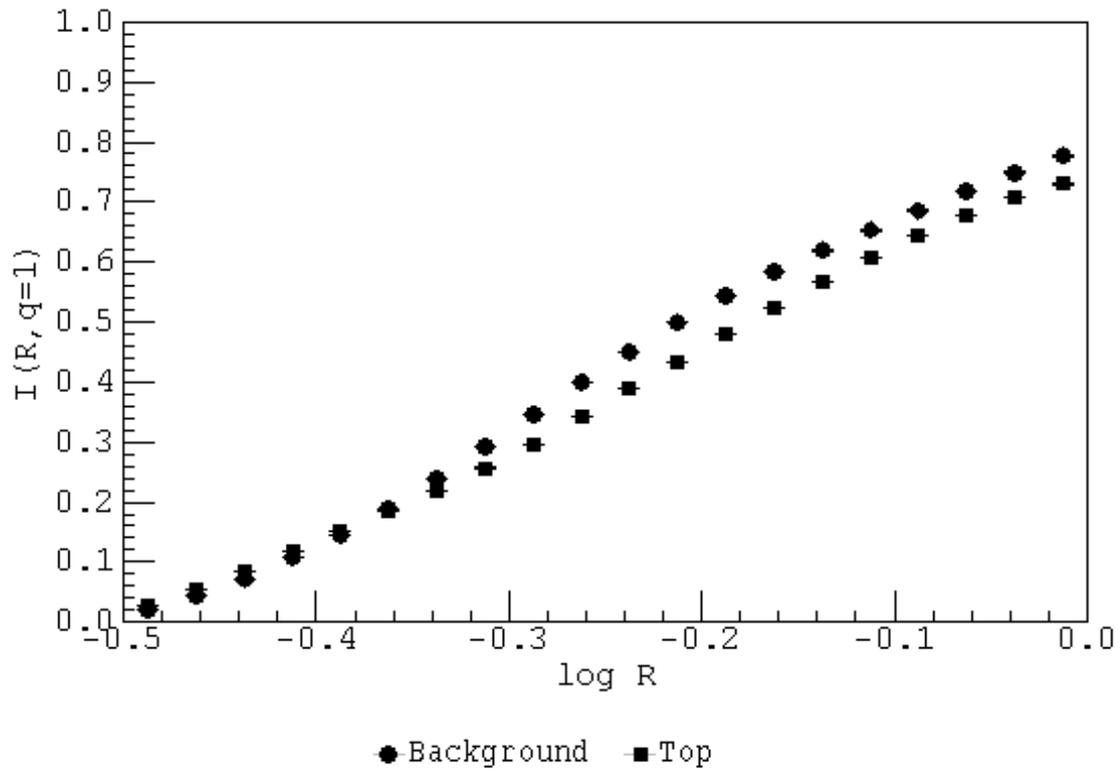

**Fig 2**. Comparison of *I(R,q=1)*'s, with *R* defined as in eq. 5, for $t\bar{t} \to$ non-leptonic and background channels.

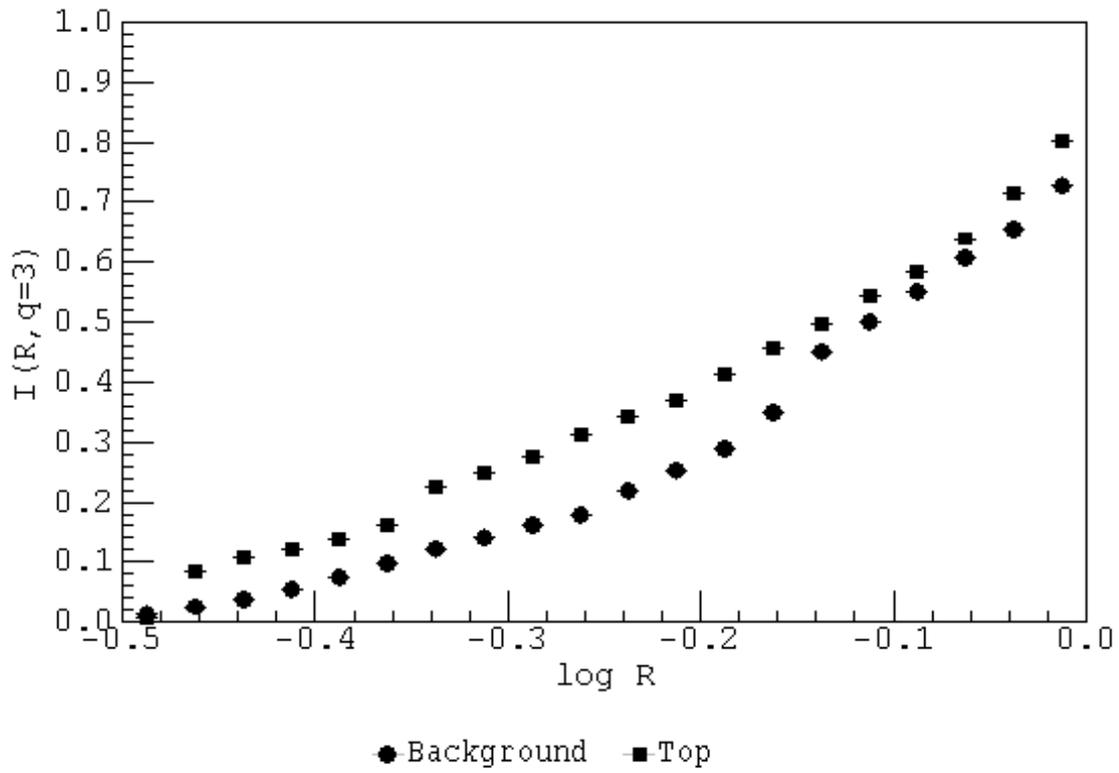

**Fig 3**. Comparison of *I(R,q=3)*'s, with *R* defined as in eq. 4, for $t\bar{t} \rightarrow$ non-leptonic and background channels.

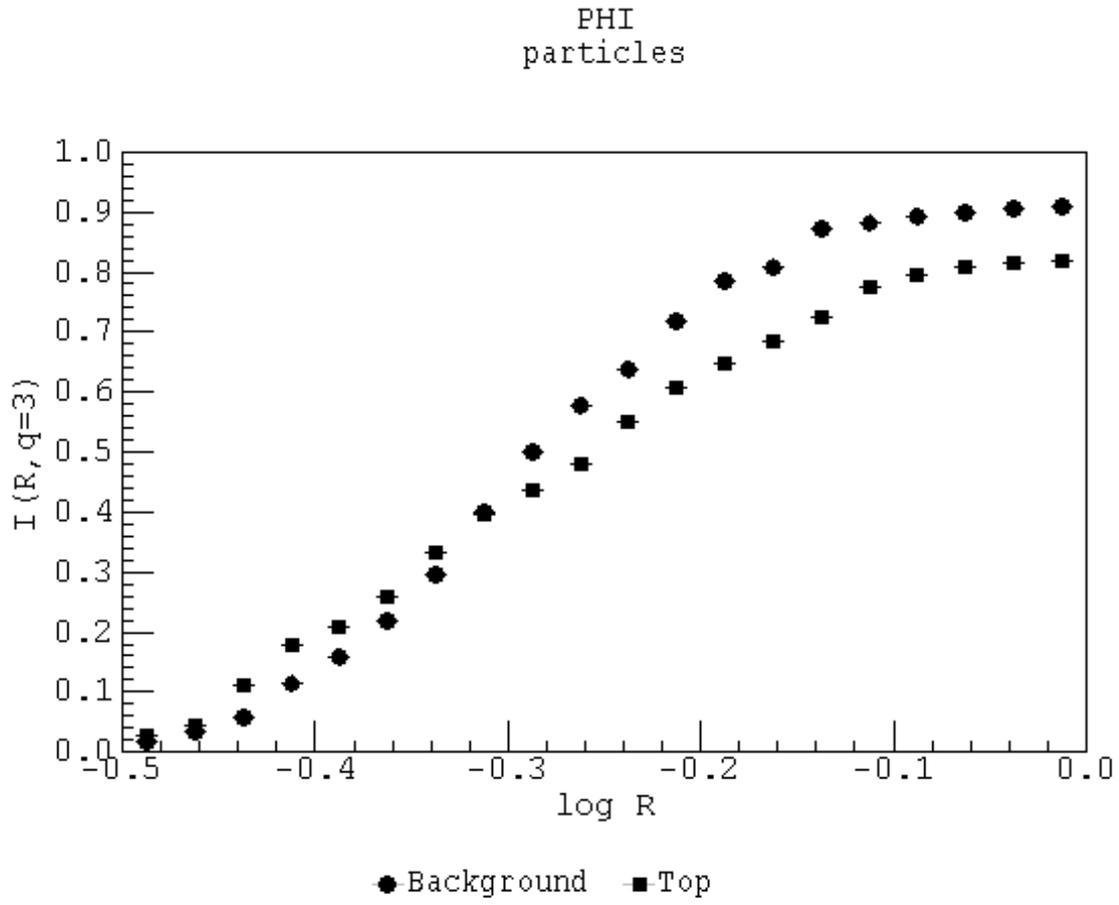

**Fig 4**. Comparison of *I(R,q=3)*'s, with *R* defined as in eq. 5, for $t\bar{t} \to$ non-leptonic and background channels.